\documentclass{article}
\usepackage{fortschritte}
\usepackage[textwidth=16cm, textheight=23.8cm]{geometry}
\usepackage{amsmath}
\usepackage{hyperref}
\usepackage{xspace}
\usepackage{graphicx}
\usepackage{psfrag}
\usepackage[numbers,sort&compress]{natbib}

\newcommand{\adss}{$\text{AdS}_5\times S^5$\xspace}

\DeclareMathOperator{\Tr}{Tr}
\newcommand{\eqn}[1]{(\ref{#1})}  

\begin{document}
{~}\vskip -4ex
\def\titleline{Splitting strings and chains}
\def\email_speaker{{\tt marija.zamaklar@aei.mpg.de}}
\def\authors{Kasper Peeters, Jan Plefka and Marija Zamaklar\hskip-.2em\sp}
\def\addresses{Max-Planck-Institute for Gravitational Physics
(Albert-Einstein-Institute),\\ Am M{\"u}hlenberg 1, 14476 Golm, Germany}

\def\abstracttext{We review a study of the semiclassical decay of
macroscopic spinning strings in \adss as well as its dual gauge
theory description. The conservation of the infinite tower of
commuting charges in the semiclassical string $\sigma$-model
description of the process suggests that the decay channel 
of maximal probability should preserve integrability in the gauge theory.}

\large
\makefront
\section{Introduction }

Probing the AdS/CFT correspondence in the regime where 
the string coupling constant $g_s$ is non-vanishing
is obviously a relevant task. It identifies string splitting and
joining interactions
with non-planar diagrams in the dual gauge theory. In recent
studies of the correspondence it has proved very fruitful to consider
a limit of large quantum numbers in both theories, enabling detailed
comparisons. The celebrated Berenstein, Maldacena and Nastase
limit~\cite{Berenstein:2002jq} considers the sector in which one
angular momentum $J_1$ on the five sphere becomes large. Here quantitative
control on the interacting string sector is available
\cite{Constable:2002hw}.

In~\cite{Peeters:2004pt} we addressed the question whether this
control over the non-planar gauge theory/interacting string sector
could be extended to the situation where two angular momenta on the
five sphere $J_1$ and $J_2$ become large. In the free string 
situation this limit corresponds to
large, macroscopic spinning strings in \adss.  The energies of these
strings and the anomalous dimensions of the dual gauge theory operators
agree in leading loop orders for the planar gauge theory, but 
not much has been done in th non-planar sector.

\begin{figure}[t]
\begin{center}
\hbox{\includegraphics*[width=.4\textwidth, angle=-90]{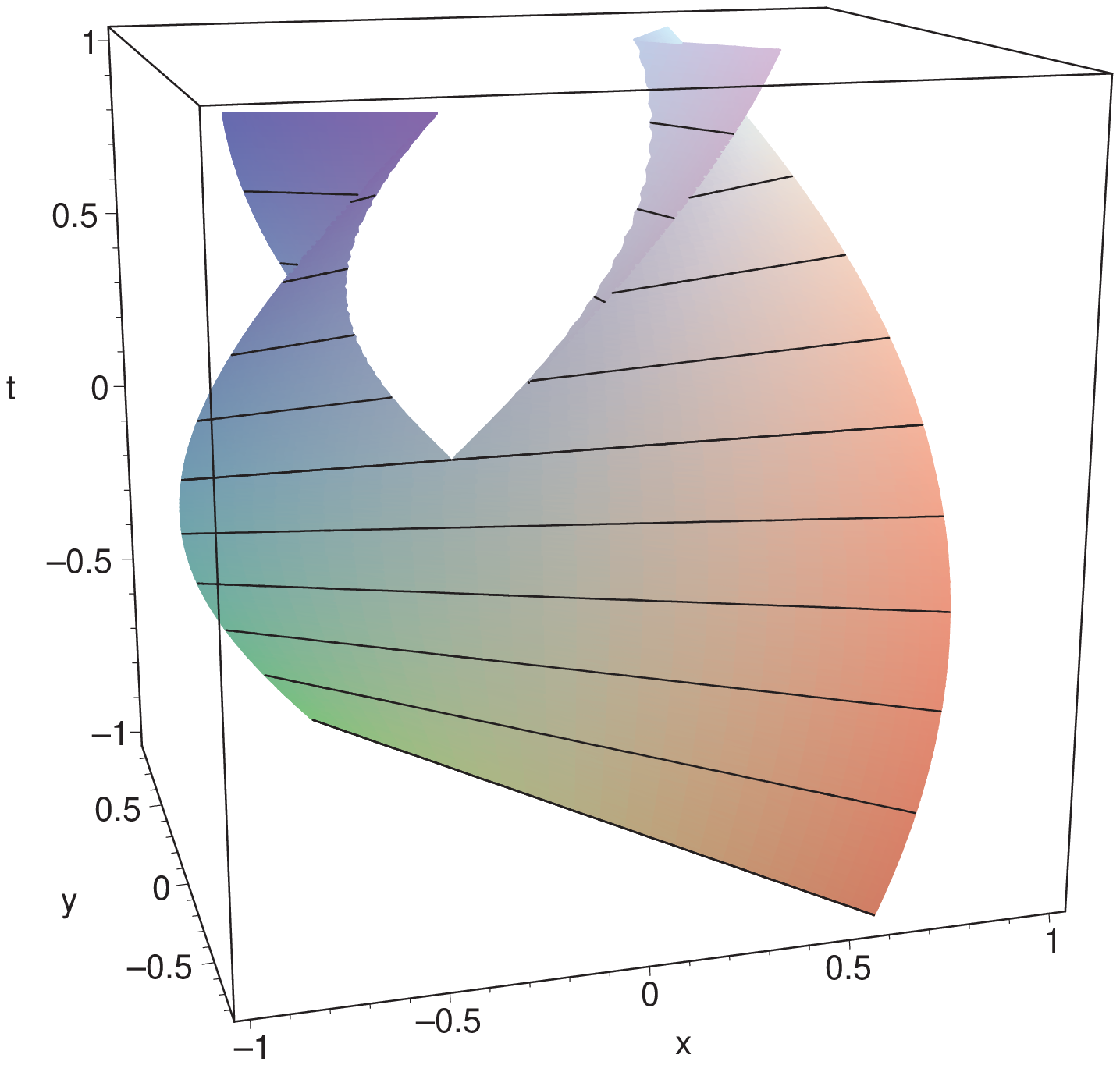}\hskip-4em
      \includegraphics*[width=.4\textwidth, angle=-90]{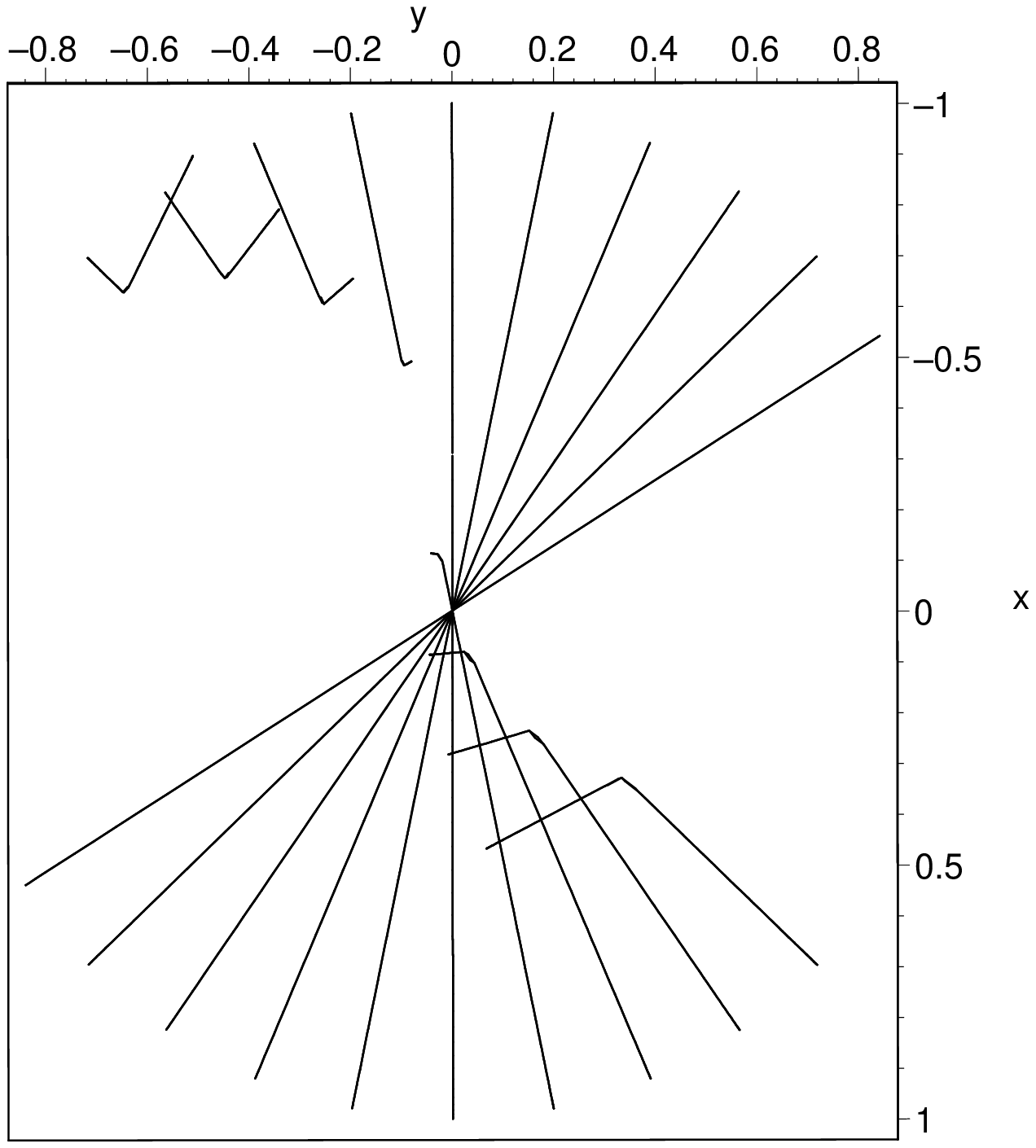}}
\end{center}
\caption{Semi-classical decay of a folded, rotating string in flat
  space-time, following~\cite{Iengo:2003ct}. The plot on the right
  shows snapshots at various values of $\tau$. The outgoing pieces
  exhibit kinks, which propagate outward along the strings. New
  momenta $P_x^I= -P_x^{II}$ are generated in the decay process.}
\label{f:splitflat}
\end{figure}

The central question addressed in our work~\cite{Peeters:2004pt} is
what can be said about \mbox{$g_s \neq 0$} effects for large spinning
strings. Although the quantum computation on \adss cannot be done at
present, it is possible to analyze the decay semi-classically. In
\emph{flat} space-time, the semi-classical decay of macroscopic
strings was analyzed in detail by Iengo and
Russo~\cite{Iengo:2003ct}. In the semi-classical approach, one starts
with a classical, rotating closed string solution. At a given
time~$\tau=0$, the string can spontaneously split if two
points~$\sigma$ and $\sigma'$ on the string coincide in target space,
and if their velocities agree. The string described by these boundary
conditions,~$X^\mu(\tau,\sigma)=X^\mu(\tau,\sigma')$ and $\dot
X^\mu(\tau,\sigma)=\dot X^\mu(\tau,\sigma')$, then forms a ``figure
eight''. The splitting is realized by declaring that from $\tau=0$
onward, each of the two string pieces (``left and right'' from the
overlapping point), \emph{separately} satisfy periodic boundary
conditions. The initial conditions on the positions and velocities of
the outgoing pieces are simply taken to be those of the incoming
string at the moment of splitting.  The effect of the splitting
propagates with the speed of light along the outgoing pieces, leading
to kink-like shapes (see figure~\ref{f:splitflat}).

\begin{figure}[t]
\begin{center}
\includegraphics*[width=0.5\textwidth]{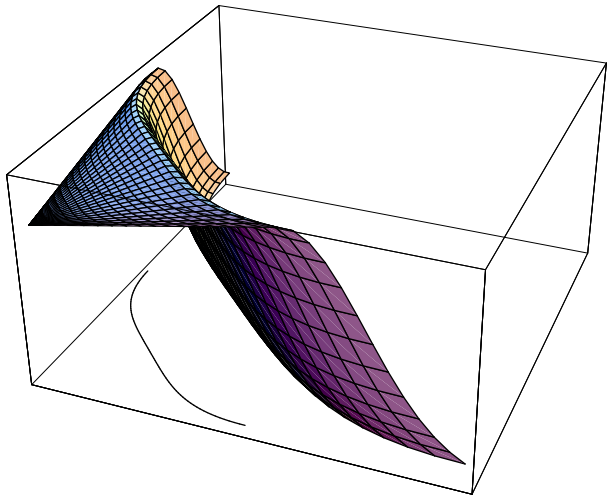}
\caption{Sketch of the relation between the semi-classical and the
  full quantum calculations. The surface depicts the quantum decay
  amplitude over the (horizontal) plane spanned by the mass-square of
  the two outgoing strings, $(M_I)^2$ and $(M_{II})^2$. The amplitude
  reaches its maximum over the curve allowed by semi-classical decay.}
\label{f:classquant}
\end{center}
\end{figure}

The relations between the energies and angular momenta of the outgoing
strings are determined completely by conservation laws, i.e.~one does
not need to derive the explicit string shapes in order to obtain these
relations. From the relations between the charges one can then produce
a curve in, for instance, the plane spanned by the masses $M_I$ and
$M_{II}$ of the outgoing string pieces. In flat space-time, this curve
can be compared with a \emph{full quantum} string computation of the
decay rate. It has been shown that
the quantum decay rate, as a function of the outgoing masses, reaches
its maximum very close to the curve obtained from the classical
analysis (see figure~\ref{f:classquant}).  

In the present paper we will review the analysis of the decay of
semi-classical strings on \adss presented in~\cite{Peeters:2004pt}. The
goal is to produce predictions which can in principle be verified on
the gauge theory side. We will focus on the folded string which is
rotating on the~$S^5$ factor of the background~\cite{Frolov:2003xy}.

\section{String splitting in AdS/CFT}

Let us consider the spontaneous splitting of the solution of
Frolov and Tseytlin~\cite{Frolov:2003xy}.  The two-spin string
solution is given by the equations
\begin{equation}
\label{e:FT2}
t = \kappa \tau \, , \quad \rho =0 \, , \quad \gamma = \frac{\pi}{2} \, , 
\quad \varphi_3 =0 \, , \quad \varphi_1 = w_1 \tau \, , \quad \varphi_2 = w_2 
\tau \,, \quad \psi = \psi(\sigma) \, , 
\end{equation}
where $\kappa,w_1$ and $w_2$ are constants. The equation which
determines the profile of $\psi(\sigma)$ is
\begin{equation}
\label{e:eoms}
 \psi^{'2}  = w_{21}^2 (\sin^2 \psi_0 - \sin^2 \psi)\,, \quad w_{21}^2 \equiv w_2^2 
- w_1^2 \geq 0 \, . 
\end{equation}
Here the constant $\psi_0$ corresponds to the target-space length of
the folded string.  The charges carried by the string are given by
\begin{equation}
{\cal E} = \int_{0}^{2 \pi}  {{\rm d} \sigma \over 2 \pi} \dot{X}^0\,,\quad
{\cal J}_{ij} =  \int_0^{2\pi} {{\rm d} \sigma \over 2 \pi} (X_i \dot{X}_j 
-  X_j \dot{X}_i)  \, , \quad (i,j=1\cdots 4) \, .
\end{equation}
Before the decay, these charges evaluate to
\begin{equation}
\label{e:J12bJ34b}
{\cal E} = \sqrt{w_2^2 \sin^2 \psi_0 + w_1^2 \cos^2 \psi_0} \,,\quad
{\cal J}_{12} = \frac{2\omega_1}{\pi\, \omega_{21}} E(q) \,,\quad
{\cal J}_{34} = \frac{2\omega_2}{\pi\, \omega_{21}} (K(q)-E(q)) \, , 
\end{equation}
where we defined $q\equiv \sin^2 \psi_0$.  The parameters $\omega_1$
and $\omega_2$ have no analogue on the gauge theory side, but they can
be eliminated completely, producing a relation between the physical
quantities, 
\begin{equation}
{\cal E} = {\cal J}\, {\cal E}_0(\alpha) + \frac{{\cal E}_1(\alpha)}{{\cal J}} + \frac{{\cal E}_2(\alpha)}
{{\cal J}^3} + \ldots\, ,
\end{equation}
where the coefficients ${\cal E}_i$ are explicitly computable functions,
which depend on the filling fraction~$\alpha= {\cal J}_{34}/{\cal J}$.
Let us now consider the splitting process. We choose a
parameterization on the world-sheet of the string which is depicted in
the figure below,
\begin{center}
\psfrag{ps0}{\small $\psi(0)=-\psi_0$}
\psfrag{ps2p}{\small $\psi(2\pi)$}
\psfrag{psap}{\small $\psi(a\pi)=\tilde\psi$}
\psfrag{ps-ap}{\small $\psi(-a\pi)$}
\psfrag{psp2}{\small $\psi(\pi)=\psi_0$}
\psfrag{cut}{\small cut}
\vspace{1ex}
\includegraphics*[width=.6\textwidth]{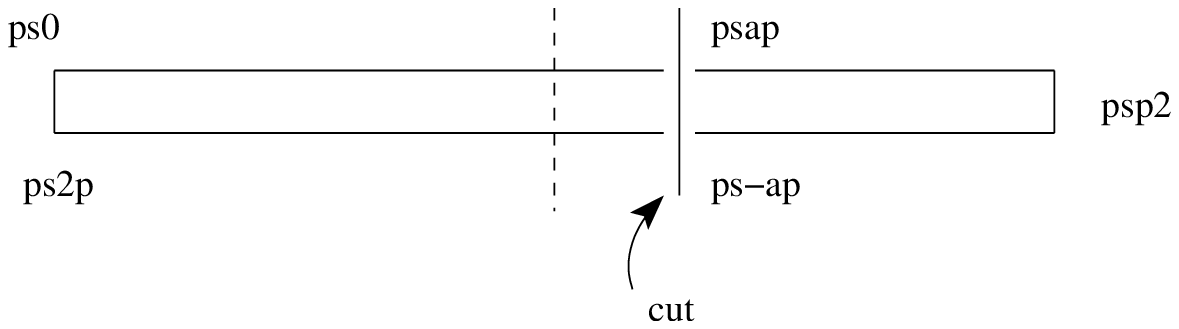}
\end{center}
The charges of the strings \emph{after} the decay can be obtained by
using the solution of the string \emph{before} the decay, but
integrating the charge densities over the lengths of each piece of
string separately (i.e.~$\sigma\in [-\pi a, \pi a]$ for the first
piece and $\sigma \in [-\pi,-\pi a] \cup [\pi a,\pi]$ for the second
one).  This is consistent, as the initial conditions generated from
the unsplit solution for the outgoing two string pieces are
consistent, i.e.~obey the Virasoro constraint.

The energies get distributed simply according to
\begin{equation}
\label{e:enIII}
{\cal E}^I =\kappa\, a \,,\qquad  {\cal E}^{II}=\kappa\, (1-a) \,,
\end{equation}
while the angular momenta $J_{12}$ and $J_{34}$ get distributed
between the outgoing string pieces~$I$ and $II$ according to (similar
expressions hold for piece~$II$; see~\cite{Peeters:2004pt})
\begin{equation}
\label{e:J12IJ34I}
{\cal J}_{12}^I = \frac{\omega_1}{\pi\omega_{21}}\left ( E(q) + E(x;q) \right )\,,\qquad
{\cal J}_{34}^I = \frac{\omega_2}{\pi\omega_{21}}\left ( K(q)-E(q) + F(x;q) - E(x;q) \right )
\end{equation}
where $x:=\arcsin(\frac{\sin\tilde\psi}{\sin\psi_0})$.
Some of the remaining angular momenta which vanish before the split
now become non-zero for the outgoing strings,
\begin{align}
{\cal J}_{14}^{I} & = - {\cal J}_{14}^{II} =
 - { w_2 \over \pi w_{21}} \sqrt{\sin^2 \psi_0 - \sin^2 \tilde{\psi}} \, , \quad
{\cal J}_{23}^{I}  =  -{\cal J}_{23}^{II} =
{ w_1 \over \pi w_{21}} \sqrt{\sin^2 \psi_0 - \sin^2 \tilde{\psi}} \, .
\end{align}
The sum of each of these momenta is zero in accordance with the
conservation laws.

\section{Invariant physical data}

The goal now is to eliminate the parameters $x$ and $q$ related to the
splitting point and initial string length, and express all conserved
charges in terms of a minimal set of independent ones. The split
introduces only one extra free parameter, namely the point~$x$ at
which the string splits, while the number of measurable charges
doubles: $\alpha^I$,$\alpha^{II}$, ${\cal J}^I$ and ${\cal
J}^{II}$. Hence after the split, the number of dependent quantities,
as well as the number of functional relations between them (which
should be compared to the gauge theory) is larger. 

\begin{figure}[t]
\hspace{-1em}\mbox{\includegraphics*[width=.5\textwidth]{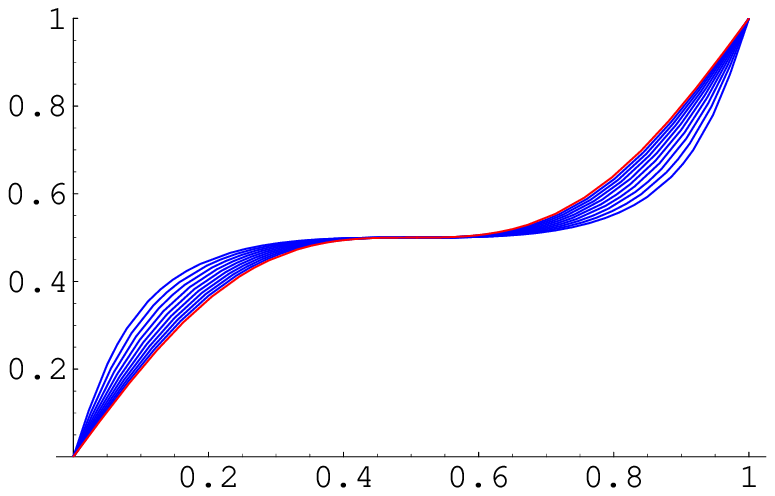}\quad
\includegraphics*[width=.5\textwidth]{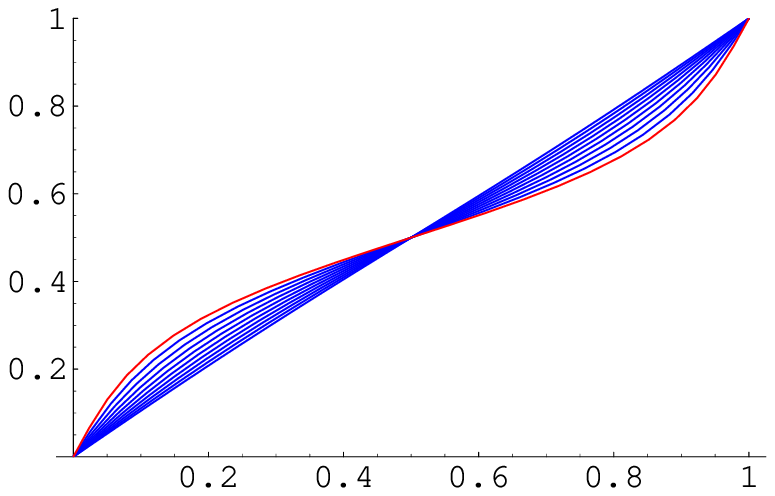}}
\caption{Figure on the left: plot of the relation between $\beta_{12}$
  (horizontal) and $\beta^0_{34}$ (vertical) as defined
  in~\protect\eqn{e:beta1234def}. The various curves correspond to
  various values for the filling fraction $\alpha\in[0.05,\ldots
  0.5]$. Note the symmetry with respect to the point $(0.5,0.5)$ as a
  consequence of the geometry of the folded string ($\psi(\sigma)=
  \psi(\pi- \sigma)$). The point $(1,1)$ corresponds to the unsplit
  string.  Figure on the right: the energy ${\cal E}_0^I$ of the first
  outgoing string as a function of $\beta_{12}$. The straight line
  corresponds to $\alpha=0.5$.\label{f:spins-filling}}
\end{figure}

The first functional relation we want to establish is the relation
between the two angular momenta carried by the first part of the string,
\begin{equation}
\label{e:beta1234def}
\beta_{12}:= \frac{{\cal J}_{12}^{I}}{{\cal J}_{12}} \,,\qquad
\beta_{34}:= \frac{{\cal J}_{34}^{I}}{{\cal J}_{34}} \,,\qquad\text{with}\qquad
{\cal J} := \underbrace{{\cal J}_{12}^I +{\cal J}_{12}^{II}}_{=:{\cal J}_{12}} + 
\underbrace{{\cal J}_{34}^I +{\cal J}_{34}^{II}}_{=:{\cal J}_{34}}  \, .
\end{equation}
Combining equations~\eqn{e:J12IJ34I} with
equations~\eqn{e:J12bJ34b} one deduces that
\begin{equation}
\label{e:beta12and34}
\beta_{12}  =  {1\over 2} \bigg( 1 + {E(x;q) \over E(q)} \bigg) \,,\quad
\beta_{34}  = {1\over 2} \bigg(1 +{F(x;q) - E(x,q) \over K(q) - E(q)} \bigg) \, .
\end{equation}
The parameter~$q$ appearing in these equations is determined by the
unsplit string. However, the splitting point~$x$ should now be
eliminated by a combination of global charges of the outgoing strings.
We decided to choose $\beta_{12}$ as the new free physical parameter
of the splitting process. Using an expansion of~$x$ in~$1/{\cal J}^2$,
\begin{equation}
\label{e:xexp}
x =  x_0 + \frac{x_1}{{\cal J}^2} + \frac{x_2}{{\cal J}^4} + \ldots \, ,
\end{equation}
one can find the coefficients $\beta_{12}(x_0, q_0)$ and $x_1(x_0,
q_0, q_1)$.  Substituting the expansion for $q$ and $x$ in the second
equation of~\eqn{e:beta12and34}, one is left with 
the functional relation $\beta_{34} = \beta_{34}(\beta_{12},
\alpha, {\cal J})$, given as a series in~$1/{\cal J}$. See
figure~\ref{f:spins-filling}a.

One might wonder whether from the gauge-theory perspective it makes
sense for the splitting parameter $x$ and the outgoing angular
momentum fraction $\beta_{34}$ to be dependent on~${\cal J}$.  After
all, the splitting Hamiltonian commutes with the R-charge
operators~${\cal J}_{12}$ and~${\cal J}_{34}$. Hence, going up higher
in perturbation theory should not induce coupling-constant dependent
modifications to the R-charges of the outgoing strings.  However, the
semi-classical string calculation captures only a part (namely the
maximum) of the full quantum surface of the decay process. The
position of the maximum varies as we go higher up in perturbation
theory. At each order in perturbation theory, the most probable
outgoing string with fixed~${\cal J}_{12}^I$ is carrying a different
${\cal J}_{34}^I$. This effectively means that the maximal probability
varies with~${\cal J}$.

The second functional relation we want to obtain is a relation between
the energy of the first outgoing piece~${\cal E}^I$ and the parameters
$({\cal J}, \alpha, \beta_{12})$. Following similar steps as in the
previous case, it is easy to derive an expression for ${\cal
E}^{I/II}$,
\begin{equation}
\label{e:EI/IIexp}
{\cal E}^{I/II}= {\cal J}\,{\cal E}^{I/II}_0 +  {\cal E}^{I/II}_1\, \frac{1}{{\cal J}}  + \ldots \, .
\end{equation}
The first coefficient in the expansion is given by
\begin{equation}
{\cal E}^{I}_0 = \frac{{\cal J}^I_{12}}{{\cal J}} + \frac{{\cal J}^{I,0}_{34}}{{\cal J}} = (1- \alpha)\, \beta_{12} + \alpha\, \beta_{34}^0 \, , 
\end{equation}
and is in agreement with
the (trivial) gauge theory prediction: the two decay products (single trace operators)
have engineering dimensions $J_{12}^I$ and $J_{34}^{I,0}$.
In figure~\ref{f:spins-filling}b we plot the energy of the first string
piece as a function of $\beta_{12}$, for various filling
fractions. The coefficient at order $1/{\cal J}$ of \eqn{e:EI/IIexp}
can be obtained as well, and yields a prediction of the anomalous
dimension at one loop of the first decay product (single trace
operator) in the dual gauge theory. Further relations can be found in~\cite{Peeters:2004pt}.

Thus far we have only discussed the behavior of the string energy and
angular momenta under the decay process. However, the classical string
sigma model is known to possess an infinite number of local, conserved
and commuting charges $Q_n$ due to its integrability
\cite{Pohlmeyer:1975nb,Ogielski:1979hv,DeVega:1992xc}.
These were written down explicitly in the work
of~\cite{Arutyunov:2003rg} for the folded string solution in terms of
a generating functional.  On the other hand, one does not expect the
string sigma model to remain integrable once string interactions are
included (i.e.~when $g_s\neq 0$). This may be seen explicitly from the
dual gauge theory side: non-planar graphs break the integrability of
the planar theory.  Nevertheless it is obvious that, for the
semi-classical decay process we are studying here, the higher
charges~$Q_n$ {\it are} conserved. This conservation follows from the
same logic that was used for the calculation of the energy and angular
momenta. If the initial charges are given via a charge density
as~$Q_n=\int\!{\rm d}\sigma\, q_n(\sigma,\tau)$, then the charges of
the outgoing strings after the split are simply
\begin{equation}
\label{e:Qnsplit}
Q_n^{I} = \int_0^{2 \pi a}\!{\rm d}\sigma\, q_n(\sigma, \tau) \, , \quad Q_{n}^{II} =
Q_n - Q^{I} \, .
\end{equation}
Here one uses the charge densities $q_n(\sigma, \tau)$ {\it before}
the split. Generating functional commuting charges of the outgoing
strings have been explicitly computed in~\cite{Peeters:2004pt}.

How is this result to be reconciled with the breakdown of integrability 
at~$g_s\neq 0$? Again we need to remember that the quantum string decay 
leads to a full surface of possible decay channels, which generically will
not preserve the charges beyond~$Q_2$. A subset of channels will, however,
preserve all~$Q_n$. It is precisely this subsector which should capture
the semiclassical string decay analyzed in the previous subsections
and is expected to dominate the decay amplitude.

\section{Splitting processes in the dual gauge theory}

Let us now turn to the discussion of the splitting process in the dual
gauge theory. In the large-$N$ limit, the dilatation operator of
${\cal N}=4$ super-Yang-Mills factorizes as the product of a universal
space-time dependent factor times a combinatorial factor acting on the
fields inside composite operators. The string splitting vertex is
encoded in the non-planar piece of this dilatation operator. In the
relevant~SU(2) sector of two chiral complex scalar adjoint fields~$Z$
and~$W$ the (space-time independent part of the) dilatation operator
is known to be~\cite{Beisert:2002bb,Peeters:2004pt}
\begin{equation}
D_2=-\frac{g_{\rm YM}^2}{8\pi^2}\, \Tr [Z,W][\check Z, \check W] \, ,
\end{equation}
where $\check Z_{ab}:=\delta/\delta Z_{ba}$ is the matrix derivative.
The action of this operator can be expressed in the language of spin
chains, by considering the action of $D_2$ on two fields in an
arbitrary single trace operator $\Tr(W A Z B)$. One finds
\begin{multline}
D_2\circ \Tr(W A Z B) =\\
 \frac{g_{\rm YM}^2}{8\pi^2}\, \Tr A\, \Bigl ( \Tr(W Z B) -\Tr(Z W B) \Bigr )
+ \frac{g_{\rm YM}^2}{8\pi^2}\, \Tr B\, \Bigl ( \Tr(Z W A) -\Tr(W Z A) \Bigr )\,.
\end{multline}
The planar (nearest neighbor) contribution is obtained when~$A$ is the
identity operator, leading to the Heisenberg XXX${}_{1/2}$
model~\cite{Minahan:2002ve}, and the remaining part forms the
splitting Hamiltonian,
\begin{equation}
D_2^{\rm planar} = \frac{g_{\rm YM}^2\, N}{8\pi^2} \sum_{i=1}^L (\delta_{i,i+1} - P_{i,i+1})\,,\qquad
D_2^{\rm splitting} = \frac{g_{\rm YM}^2}{8\pi^2} \sum_{i,j}
(\delta_{i,j} - P_{i,j})\, {\cal S}_{ij}\,,
\end{equation}
with $P_{i,j}$ the permutation operator permuting the fields (spins)
at sites $i$ and $j$. The splitting operator~${\cal S}_{ij}$ acts in a somewhat
complicated way on the sites~$i$ and~$j$~\cite{Peeters:2004pt}.
That is, we have a Heisenberg exchange interaction multiplied by a
chain splitting operation.

While the dilatation operator is thus under control, the initial gauge
theory operator dual to the single folded string solution with angular
momenta ${\cal J}_{12}$ and ${\cal J}_{34}$ is less understood. The
dual gauge operator may be written as
\begin{equation}
\label{e:sketchstate}
\Tr(Z^{{\cal J}_{12}}W^{{\cal J}_{34}}) + \ldots
\end{equation}
where the dots stand for suitable permutations of the $Z$ and $W$'s --
which are of essential importance for the evaluation of decay
amplitudes!  The spin chain picture has proved to be very efficient
for the task of diagonalizing $D_2^{\rm planar}$ for long operators
(${\cal J}\to\infty)$ with the technology of the Bethe ansatz. This
technology allows one to find energy eigenvalues,
\begin{equation}
\label{e:BetheEnergy}
D_2^{\rm planar}\, |\psi\rangle = 
\frac{g_{\rm YM}^2\, N}{2\pi^2} \sum_{i=1}^{{\cal J}_{34}} \sin^2\left(\frac{p_i}{2}\right)\,
|\psi\rangle \, .
\end{equation}
where $p_i$ are the quasi-momenta. For this problem, one does not have
to write down the eigenstate. For the splitting process, however, one
would need this state explicitly. Denote by $|\{m_1,m_2,\ldots,
m_{J_{34}}\}\rangle_{L}$ the single trace operator of length $L$ with
$W$'s appearing at positions $m_i$. The eigenstate is
then~\cite{Bethe:1931hc},
\begin{align}
\label{e:BetheState}
|\psi\rangle =\qquad\qquad &\\ \sum_{\substack{
1\leq m_1<m_2<\ldots \\[1ex]
\ldots <m_{{\cal J}_{34}} \leq L}}&\;\;
\sum_{{\cal P}\in {\rm Perm}_{{\cal J}_{34}}} \exp\Bigl [{i\sum_{i=1}^{{\cal J}_{34}} p_{{\cal P}(i)}
\cdot m_i + \frac{i}{2}\sum_{i<j}^{{\cal J}_{34}}\varphi_{{\cal P}(i),{\cal P}(j)}}\Bigr ]\,
\Bigl |\{m_1,m_2,\ldots,m_{{\cal J}_{34}}\}\Bigr \rangle_{L} \nonumber
\end{align}
$\varphi_{ij}$ are the scattering phases respectively, and the second
sum is over all ${\cal J}_{34}!$ permutations of the labels
$\{1,2,3,\ldots, {\cal J}_{34}\}$.  In order to make contact to our
semiclassical string considerations we need to take the thermodynamic
limit $L,{\cal J}_{34}\to \infty$ with ${\cal J}_{34}/L=\alpha$
fixed. Due to the unknown structure of the continuum limit of the
permutation group the Bethe wave function (not to mention the action
of the splitting Hamiltonian) becomes a monstrous object in this
limit.  This is in stark contrast to the Bethe equations, which
actually simplify in the same limit.  This is the core of the problem
which hampers a direct analytic computation of the splitting in the
gauge theory.  In principle one could attempt to address this problem
numerically. Here however, one faces technical limitations, as the
minimal length of the spin chain for which distinguishable structures
limiting to the continuum folded string configuration start to emerge
is~26 (with half filling fraction) \cite{Beisert:2003xu}.  The
corresponding wave function~$|\psi\rangle$ contains roughly~$4\cdot
10^5$ terms, many of which have coefficients of the same order.

There are two key properties which one \emph{can} verify in the dual
gauge theory, or examine in some detail in certain toy calculations
which exemplify the general logic of the quantum
decay~\cite{Peeters:2004pt}. The first one concerns the SU(2)
structure of the decay products.  This symmetry is realized through
the operators
\begin{equation}
\label{su(2)op}
J_z \equiv J_{12 } - J_{34} =\Tr(W\check W- Z\check Z)\, , \qquad J_+=\Tr(W\check Z)\, ,\qquad J_-=\Tr(Z\check W)\, .
\end{equation}
The total spin and the $J_z$ charge of a given initial state is
conserved in the decay process. However, a highest-weight state will
generically not decay into the product of two highest-weight states:
from the semi-classical calculation we see that the decay products are
not rigid, but turn on an infinite number of modes. The dual statement
is that the ``decay products'' in the gauge theory are no longer
highest-weight states. The second property concerns the higher local
charges of the Heisenberg XXX${}_{1/2}$
chain~\cite{Beisert:2003tq}. These higher charges are generically not
preserved in this decay process. However, in the thermodynamic limit
we expect the decay to be dominated by the channels which \emph{do}
preserve all higher charges.
\medskip

\section{Outlook}

We have reviewed the computation of the semi-classical decay of
strings in \adss and the formalism for the dual gauge theory
computation~\cite{Peeters:2004pt}.  The complexity of the Bethe wave
function is the main obstacle against making a direct comparison. One
possible simplification can perhaps be obtained by using the coherent
state wave function. However, a potential
problem in this approach seems to arise from the inability to write
down wave functions for the outgoing strings. An additional guideline
for a better analytic understanding is the existence of the higher
local charges.  The decay channels in which these charges are
conserved are expected to correspond to semi-classical decay, and form
only a small subsector of all possible channels.
\vfill

\begin{small}
\begingroup\raggedright\endgroup
\end{small}

\end{document}